    \documentclass[12pt]{article} 
    \usepackage{psfig}  
    \usepackage{epsfig}  
    \usepackage{graphics}
     \newlength{\dinwidth}                       
     \newlength{\dinmargin}                      
\def\Journal#1#2#3#4{{#1} {\bf #2}, #3 (#4)}

\def\NPB{{\em Nucl. Phys.} B}
\def\PLB{{\em Phys. Lett.}  B}

\def\PRD{{\em Phys. Rev.} D}
\def\ZPC{{\em Z. Phys.} C}
\def\CPC{\em Comp. Phys. Comm.}  
\def\PR{\em Phys. Rep.}          
\def\lsim{\mathrel{\rlap{\lower4pt\hbox{\hskip1pt$\sim$}}
    \raise1pt\hbox{$<$}}}                
\def\gsim{\mathrel{\rlap{\lower4pt\hbox{\hskip1pt$\sim$}}
    \raise1pt\hbox{$>$}}}                

\newcommand{\ie}{{\it i.e.\ }}

\newcommand{\Pma}{I\!\!P}
\begin{document}
\noindent
TSL/ISV-99-0215    \hfill ISSN 0284-2769\\
August 1999        
\vspace*{5mm}
\begin{center}  \begin{Large} \begin{bf}
Soft Colour Exchanges and the Hadronic Final State\footnote{In proceedings 
`Monte Carlo generators for HERA physics', DESY-PROC-1999-02, 
www.desy.de/\~{}heramc}
\\
  \end{bf}  \end{Large}
  \begin{large}
A.~Edin$^a$, G.~Ingelman$^{ab}$, J.~Rathsman$^c$\\
  \end{large}
\end{center}
$^a$ Deutsches~Elektronen-Synchrotron~DESY, 
     Notkestrasse~85,~D-22603~Hamburg,~FRG\\
$^b$ Dept.\ of Radiation Sciences, Uppsala University, 
     Box 535, S-751 21 Uppsala, Sweden\\
$^c$ Stanford Linear Accelerator Center,
Stanford University, Stanford, California 94309, USA\\
\begin{quotation}
\noindent
{\bf Abstract:}
The models for soft colour interactions and colour string re-interactions,
implemented in the Monte Carlo program {\sc Lepto}, are investigated regarding 
hadronic final states in inclusive and diffractive deep inelastic scattering.
\end{quotation}

\section{Introduction}

The hadronic final state in inclusive and diffractive deep inelastic scattering
(DIS) can give a better understanding of the interplay between soft and hard
processes in QCD. Whereas hard interactions are well described by perturbative
QCD, soft interactions are not calculable within perturbation theory. Instead
more phenomenological models are used to transform the perturbative partonic 
final state into an observable hadronic final state. 
It is normally assumed that the colour topology of an event
is given by the planar approximation in perturbation theory, so that terms of
order $1/N_C^2$ are neglected, and that this topology is not altered by soft
interactions.

\begin{figure}[htbp]
\center
\epsfig{file=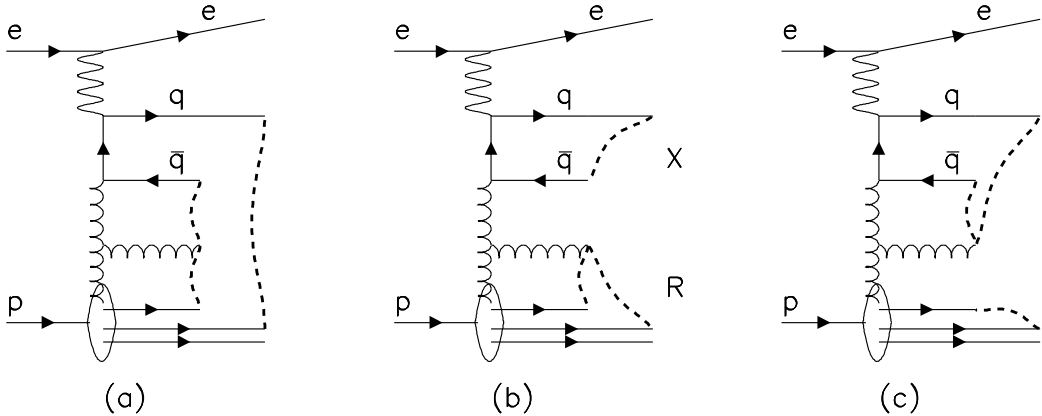,height=6cm}
\caption{{\it A gluon-induced DIS event with examples of colour string 
connection (dashed lines) of partons in 
(a) conventional Lund model based on the colour order in perturbative QCD, and 
(b,c) after colour rearrangement through SCI or GAL mechanisms.
}}
\label{fig:string}
\end{figure}

The models for soft colour interactions (SCI) \cite{SCI} and the generalised
area law (GAL) \cite{GAL} for colour string re-interactions 
try to model additional soft colour exchanges which neither belong to the
perturbative treatment nor the conventional hadronisation models. These soft
colour exchanges can alter the colour topology and thereby produce a different
final state, including such phenomena as large rapidity gaps and diffraction, 
as illustrated in Fig.~\ref{fig:string}. 

In these models there is no sharp distinction between inclusive and diffractive
events, which is the case in Regge-inspired models. Instead, there is a 
continuous transition between the different final states. The common assumption
for the two models is that the soft colour exchanges factorises from the hard
interactions which can therefore be described by standard perturbative methods,
\ie with matrix elements and parton showers. It is also assumed that compared 
to the perturbative interactions the momenta in the soft colour exchanges can 
be neglected and that their effect will be washed out by the hadronisation.

In this note we investigate the hadronic final states in inclusive and
diffractive DIS resulting from the SCI and GAL models as
implemented in the Monte Carlo program {\sc Lepto} \cite{LEPTO}. 
In section 2 we give a short review of the two models.
In section 3 we show how  the diffractive structure function can be used to
fix the amount of soft colour exchanges in the two models and compare with data
on the hadronic final state in diffractive events (the $X$-system). Section 4
then compares the two models with data on inclusive hadronic final states.
Finally, in section 5 we summarise and conclude. 

\section{Models for soft colour exchanges}

The basic assumption of the soft colour interaction (SCI) model \cite{SCI} is
that the partons produced in the hard interaction can have soft colour
exchanges with the background colour field of the incoming hadron or hadrons. 
These exchanges can change the colour topology of the event as illustrated
in Fig.~\ref{fig:string}. The probability for a soft colour exchange depends on
non-perturbative dynamics and is thus not calculable at present and for
simplicity it is therefore assumed to be a constant in the SCI model. Its
value, $R=0.5$, is obtained by comparing the model with the diffractive structure
function in DIS. As long as the SCI model represents interactions with a
colour background field, it should only be applied to reactions with initial
state hadrons.

Apart from being applicable in DIS the SCI model has also
been successfully used to describe the surprisingly large quarkonium
cross sections observed at the  Fermilab Tevatron \cite{onium}. A first
comparison with quarkonium photoproduction at HERA is presented in
\cite{heavy}. In addition the model describes diffractive $W$ and jet
production at the Tevatron \cite{Rikard,GI-DIS99,EIT}.

The generalised area law (GAL) model \cite{GAL} for colour string 
re-interactions is similar in spirit to the SCI model in that it is a model 
for soft colour exchanges. The main difference is that the GAL model is
formulated in terms of interactions between the strings connecting the partons
produced in an event. Thus the GAL model is also  applicable for hadronic final
states in $e^+e^-$, since it treats string re-interactions and should apply to
all interactions producing strings.

Another important feature of the GAL model is that the probability for an 
interaction is not constant as in the SCI model. Instead there is a dynamical 
suppression factor giving the probability $R=R_0\exp(-b\Delta A)$ for a string
reconnection, where $\Delta A$ is the difference between the areas in momentum
space spanned by the strings in the two alternative string configurations and
$b$ is one of the hadronisation parameters in the Lund model \cite{lundstring}. 

The parameters of the GAL model were obtained \cite{GAL} by making a 
simultaneous
tuning to the diffractive structure function in DIS and the charged particle 
multiplicity distribution and momentum distribution for $\pi^{\pm}$ in $e^+e^-$ 
annihilation at the $Z$-resonance. This resulted in $R_0=0.1$, $b=0.45$
GeV$^{-2}$ and $Q_0=2$ GeV, where $Q_0$ is the cut-off for initial and final
state parton showers. It is not possible to have the {\sc Jetset} default
cut-off $Q_0=1$ GeV in the parton showers and simultaneously reproduce 
the multiplicity distribution. One might worry that the obtained cut-off is 
relatively large compared to the default value. However, it is not obvious
that perturbation theory should be valid for so small scales when more
exclusive final states are considered. Therefore, $Q_0$ can be considered as a
free parameter describing the boundary below which it is more fruitful to
describe the fragmentation process in terms of strings instead of perturbative
partons.

Both the SCI and GAL models have been implemented in the LSCI routine in the
Monte Carlo program {\sc Lepto}~\cite{LEPTO}. For the GAL model one also needs
a new version of subroutine LEPTO, see the GAL homepage {\tt
http://www3.tsl.uu.se/thep/rathsman/gal} for details.

\section{Hadronic final states in diffractive DIS}

\begin{figure}[htbp]
\begin{center}
\mbox{\epsfig{figure=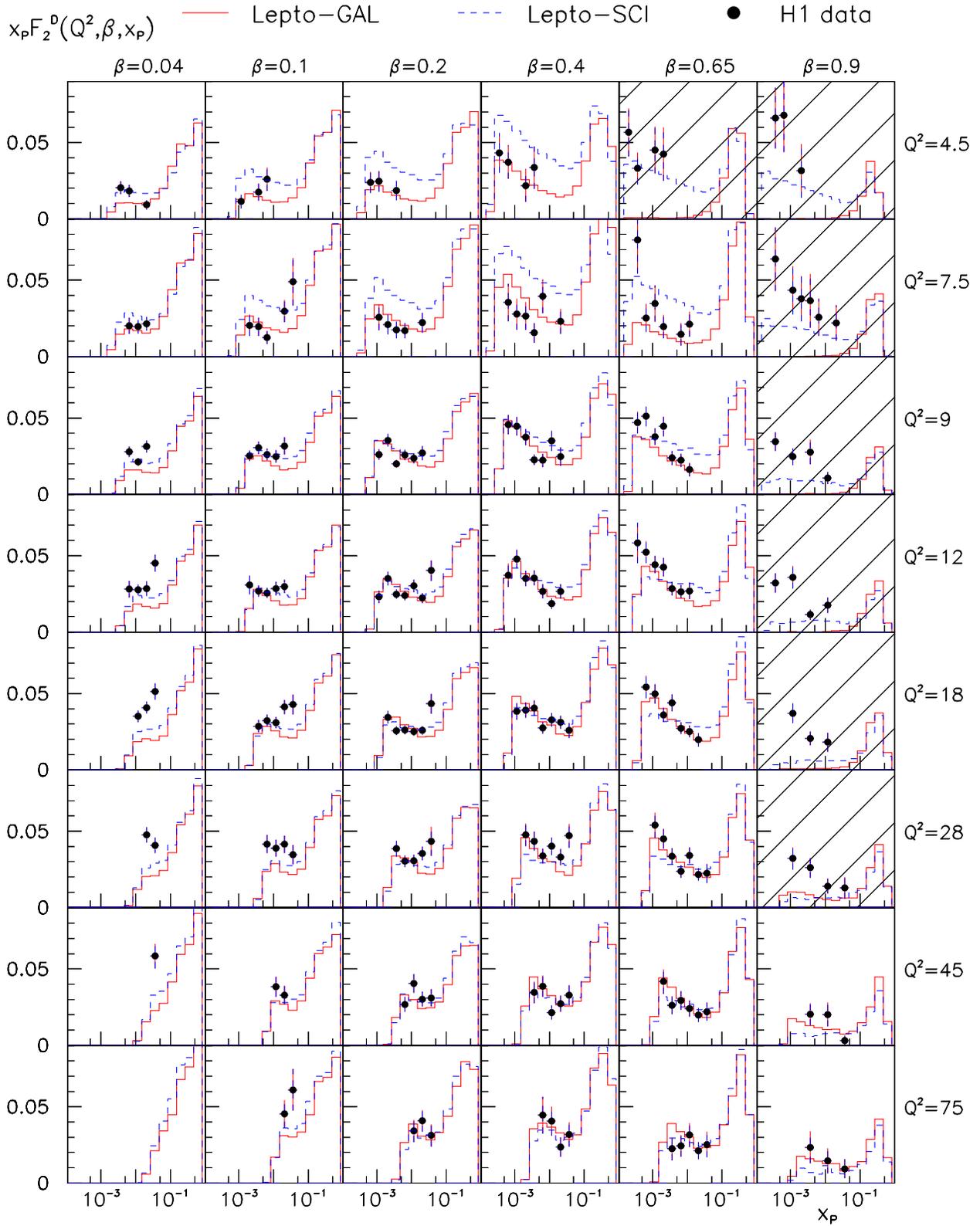,width=17cm}}
\end{center}
\vspace*{-5mm}
\caption[*]{{\it
The diffractive structure function $x_{\Pma}F_2^D$ versus $x_{\Pma}$ 
for bins in $\beta$ and $Q^2$. H1 data \cite{h1data} compared to the 
results of the GAL (full curve) and SCI (dashed) models in {\sc Lepto}. 
The hashed plots corresponds to $M_X<2$ GeV not included properly 
in the models due to the matrix element cut-off. 
}}
\label{fig:f2d3}
\end{figure}

The diffractive structure function in DIS was obtained from the SCI and GAL 
models using a subroutine from the HzTool package \cite{hztool} and the 
CTEQ4 leading order parton distributions~\cite{CTEQ4}. The results are 
compared with H1 data \cite{h1data} in Fig.~\ref{fig:f2d3}. 
The normalization parameters in the models, $R$ and $R_0$ respectively, 
were determined from this data. The default version of {\sc Lepto} was used,
except for the GAL model having the modified values of the cut-off in the 
parton showers and the hadronisation parameter $b$. In addition, 
version 2 of the sea-quark
treatment (see \cite{SCI}) was used for the GAL model with the width of the
mean virtuality set to  0.44 GeV. However, the result is not sensitive to this
choice. 

The agreement between the resulting diffractive structure function calculated 
from the two models and  H1 data is quite good as is shown in 
Fig.~\ref{fig:f2d3}, especially if one takes into account that there is only
one free parameter in the models. The variables  
$x_{\Pma}\simeq\frac{Q^2+M_X^2}{Q^2+W^2}$ and
$\beta\simeq\frac{Q^2}{Q^2+M_X^2}$ are defined in terms of observable 
invariants that do not require interpretation within a particular model.  As
usual, $Q^2$ is the photon virtuality and $W$ the mass of the complete hadronic
system. $M_X^2=Q^2\frac{1-\beta}{\beta}$ is the mass of the  diffractive system
$X$.

The Regge framework requires pomeron exchange at small $x_{\Pma}$ and other 
Regge exchanges in the transition region $0.01<x_{\Pma}<0.1$, whereas the SCI
and GAL models describes the whole region in a more economic way. The GAL
model fails only for small $M_X$ which are not included in the model because
of the cut-off $M_X^2>4$ GeV$^2$ in the matrix-element. The SCI model also
gives a good description of the data except for small $Q^2$ and small $M_X^2$.
The reason for the SCI model overshooting the data at small $Q^2$ is probably
related to the typically small number of  perturbative partons produced at
small $Q^2$. This in turn means that effectively the probability for a rapidity
gap becomes larger. In the extreme case of only four partons in the final
state the probability for a rapidity gap in the SCI model is $R=0.5$ since 
there are only two possible string configurations.

One may ask whether this kind of soft colour exchange models are essentially
models for the pomeron. This is not the case as long as no pomeron or Regge 
dynamics is introduced. The behaviour of the data on $F_2^D(\beta,Q^2)$, 
usually called the pomeron structure function, is in the SCI/GAL models 
understood as normal perturbative QCD evolution in the proton.  The rise with
$ln Q^2$ also at larger $\beta$ is simply the normal behaviour at the small
momentum fraction $x=\beta x_{\Pma}$ of the parton in the proton. Here,
$x_{\Pma}$ is only an extra variable related to the gap size or  $M_X$ which
does not require a pomeron interpretation.  The flat $\beta$-dependence of
$x_{\Pma} F_2^D=\frac{x}{\beta} F_2^D$  is due to the factor $x$ compensating
the well-known increase at small-$x$  of the proton structure function $F_2$. 
For details of this and a general review of diffractive hard scattering  see
\cite{GI-review}. 

\begin{figure}[htbp]
\begin{center}
\mbox{\epsfig{figure=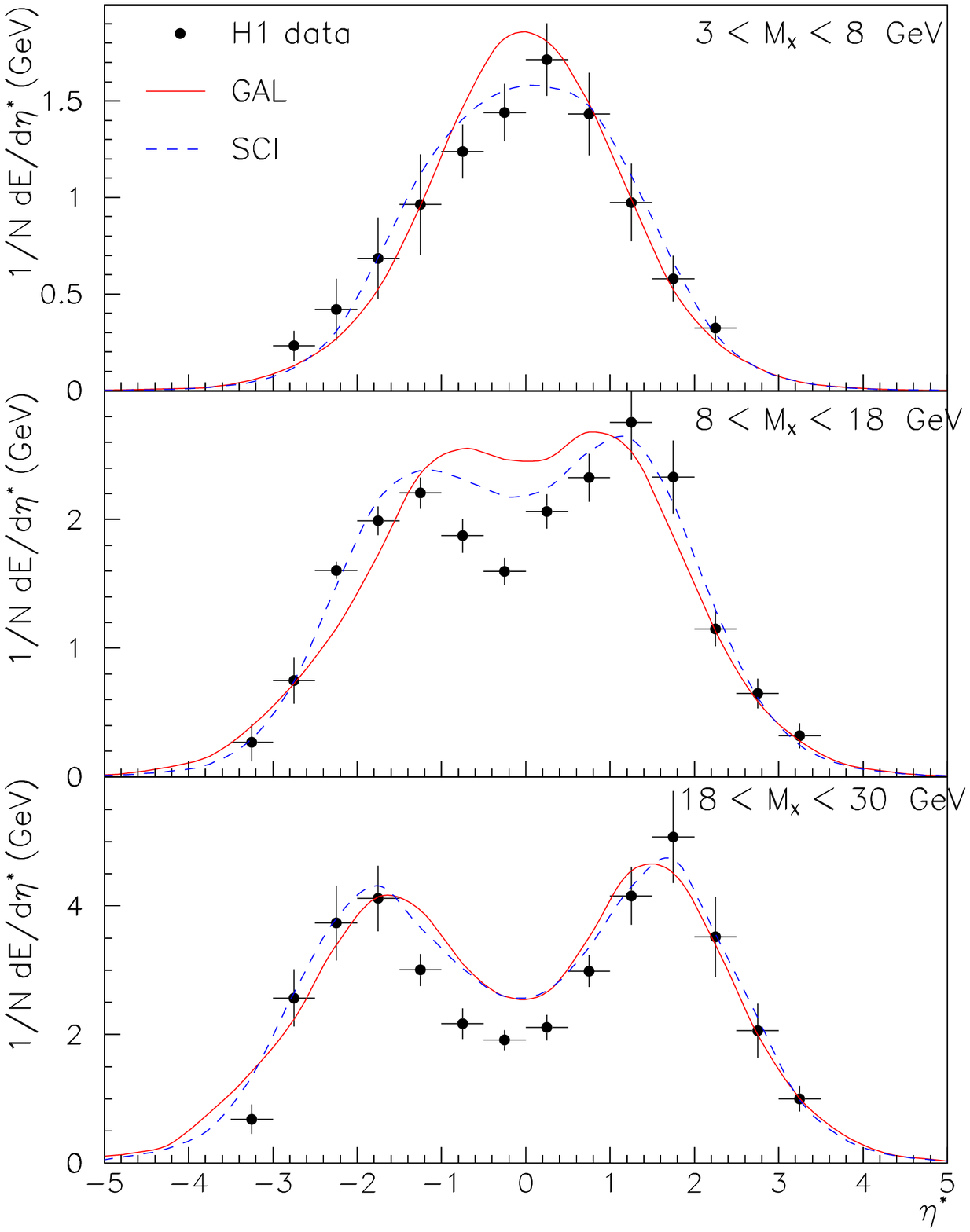,width=9cm}
\hspace*{-8mm}
\epsfig{figure=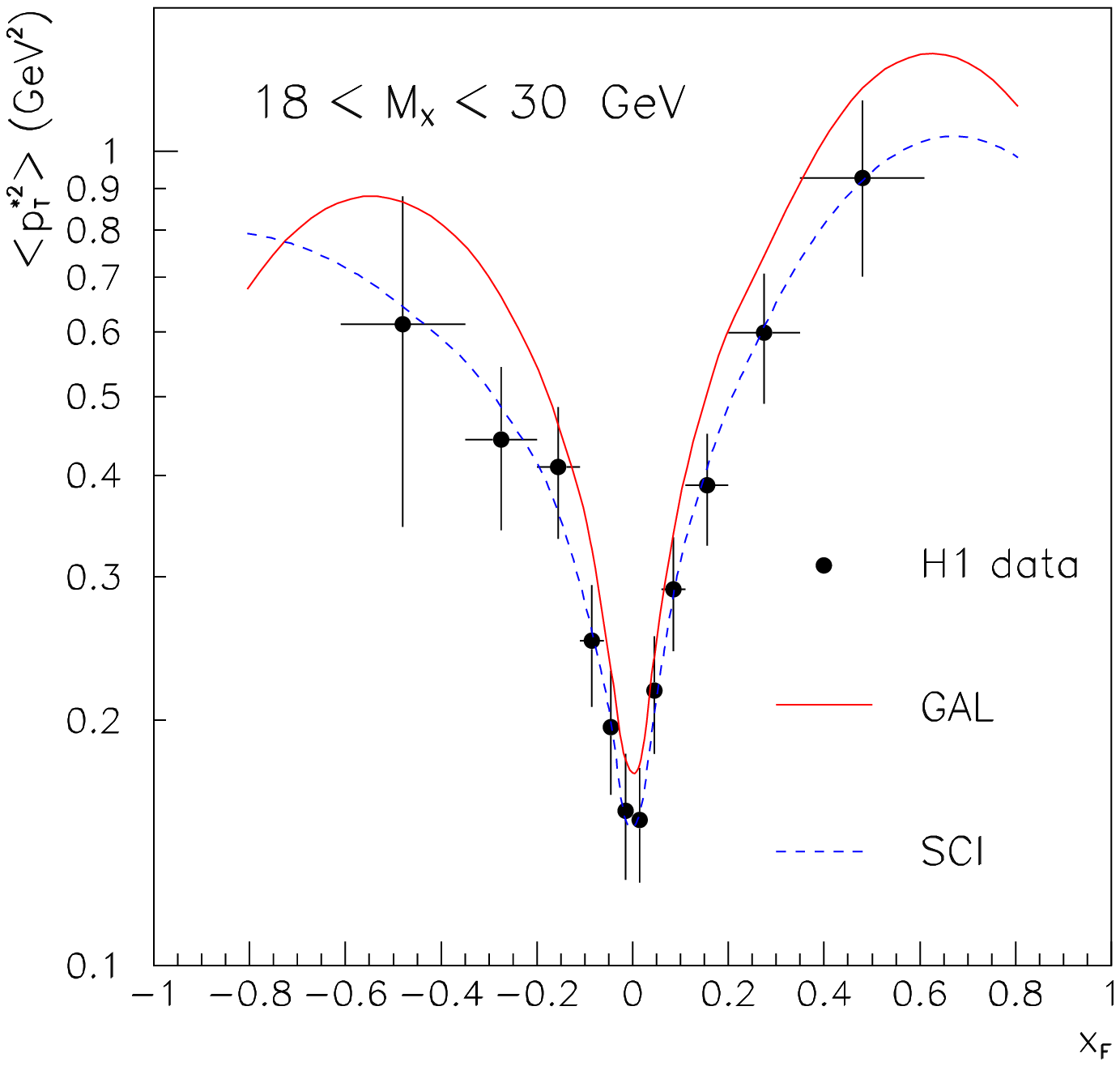,width=9.5cm}}
\end{center}
\caption[*]{{\it
(a) Energy flow versus pseudo-rapidity in diffractive H1 events
\cite{H1-diff}.
(b) Seagull plot of mean transverse momentum squared versus Feynman-$x$ 
in diffractive H1 events with $18<M_X<30$ GeV \cite{H1-diff}. 
The cms of the diffractive $X$-system is used and the curves are from 
the GAL (full) and SCI (dashed) models. Event selection:
$7.5 <Q^2 < 100$ GeV$^2$, $0.05<y<0.6$, $x_{\Pma} < 0.025$.
  }}
\label{fig:diffeflowseagull}
\end{figure}

With the free parameters of the two models fixed from the diffractive
structure  function the models can be tested by comparing with the hadronic
final state  in diffractive events. The energy flow in
Fig.~\ref{fig:diffeflowseagull}a  demonstrates that both models give a
reasonable description of the data,  with the SCI model doing slightly better.
The `seagull' plot in Fig.~\ref{fig:diffeflowseagull}b also shows that the  SCI
model is very close to data and that the GAL model gives a reasonable 
description although the transverse activity is on the high side.

There are many other observables in diffractive events to which the models 
could be compared; in particular those related to the proton remnant system, 
such as 
$t$-dependence, momentum distribution for leading protons and neutrons etc.
However, these observables are not directly related to the hadronic final state
in the $X$-system and depend on a different part of the model contained in {\sc
Lepto}. Therefore we do not study such observables here. They deserve a
dedicated investigation as initiated in \cite{SCI}.

\section{Inclusive hadronic final states}
With both models giving a good description of the hadronic final states in
diffractive events it is imperative to check that they also can describe the
inclusive hadronic final states in DIS. Energy flows in the hadronic cms is an
important observable which we have investigated earlier \cite{SCI} and H1 has
recently made a comprehensive comparison of their data with several models
\cite{H1-eflow}. However, a more detailed test is obtained by looking at the 
$p_\perp$-spectrum for charged particles which is sensitive to the distribution
of transverse energy and not only the average. We therefore consider this and 
other observables in the following. 

\begin{figure}[hbt]
\begin{center}
\mbox{\epsfig{figure=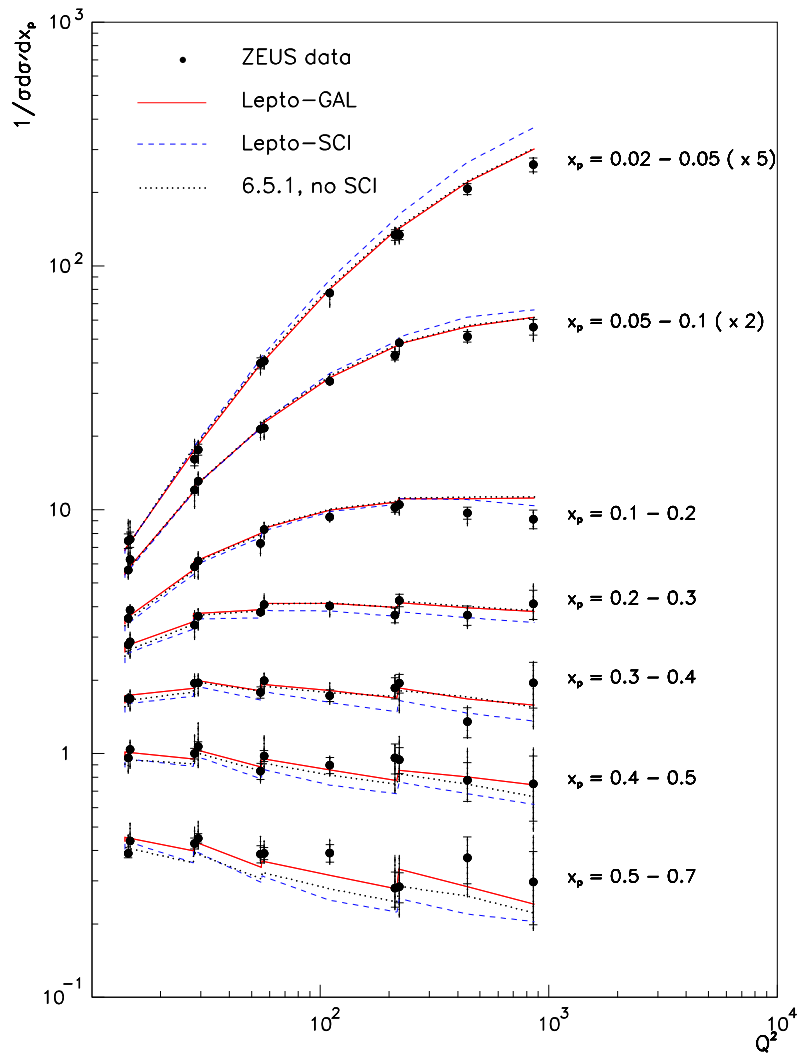,width=10.5cm}}
\end{center}
\vspace*{-10mm}
\caption[*]{{\it
$Q^2$-dependence of scaled momentum $x_p$ of hadrons in the current region
of the Breit frame in DIS.
  }}
\label{fig:mult}
\end{figure}

A good starting point for such an investigation is the momentum distribution 
of particles in the current region of the Breit frame. This part of phase-space
is expected to be well described by the models since it should not be affected 
by the proton remnant and therefore be similar to $e^+e^-$-annihilation.  The
distribution of scaled momentum $x_p=2|\bar{p}|/Q$ in this system is shown in
Fig.~\ref{fig:mult}. Although the overall agreement between the ZEUS data
\cite{ZEUS-xp} and the models is reasonable, it is clear that the SCI model
gives too many soft particles (low $x_p$) and too few hard (high $x_p$) ones.
The GAL model and also {\sc Lepto} without string topology rearrangements,
describes the details of the data quite well.

\begin{figure}[t]
\begin{center}
\mbox{\epsfig{figure=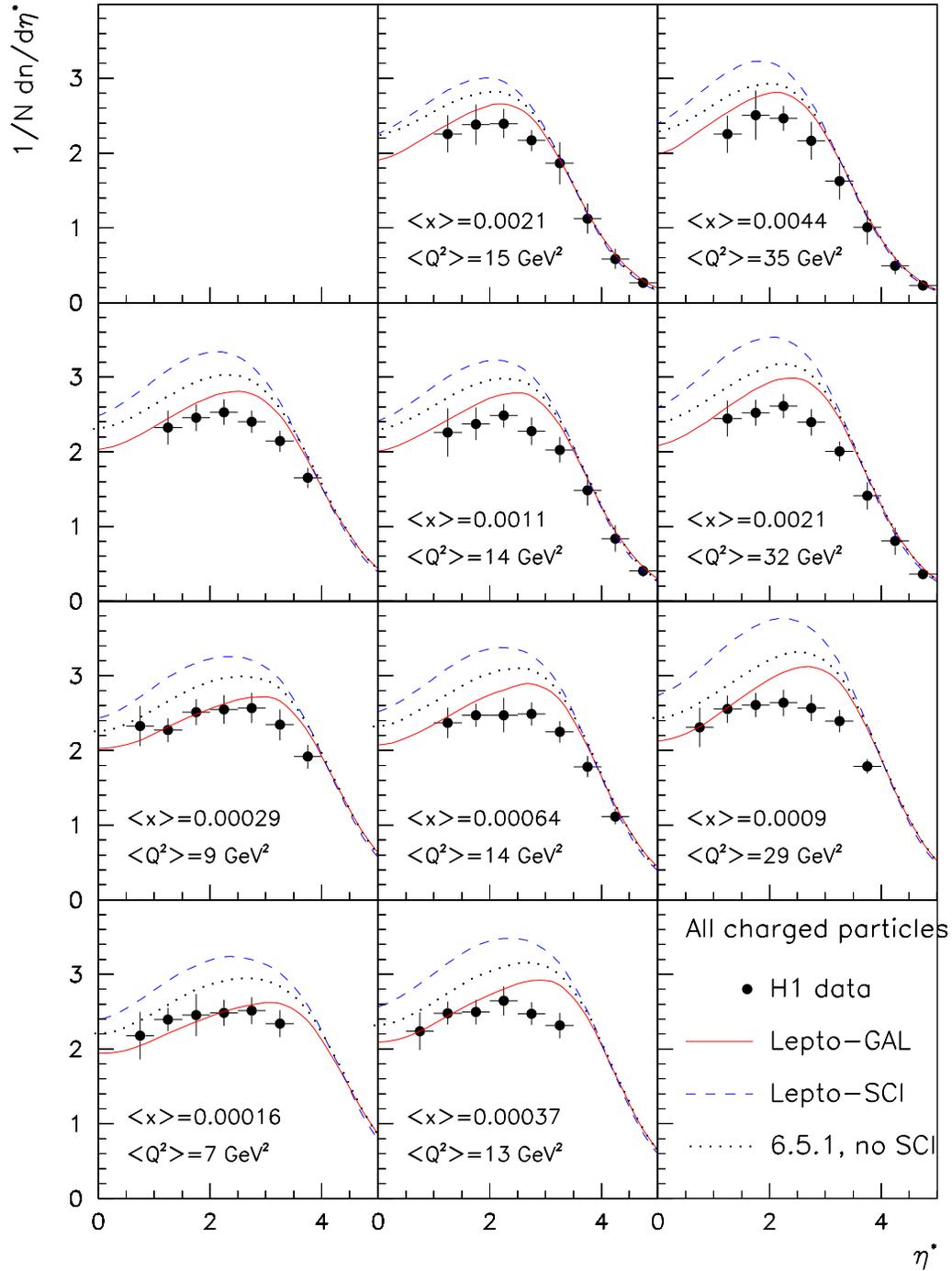,width=14cm}}
\end{center}
\vspace*{-10mm}
\caption[*]{{\it
Pseudo-rapidity distribution of charged particles in the
hadronic cms of inclusive DIS events.
  }}
\label{fig:etaspecall}
\end{figure}

The pseudo-rapidity distribution of charged particles in the detectable 
regions of the hadronic cms is shown in Fig.~\ref{fig:etaspecall}.  Again the
SCI model gives too many soft particles, whereas the GAL model is much closer
to data and even better than {\sc Lepto} without  reconnections.

\begin{figure}[t]
\begin{center}
\mbox{\epsfig{figure=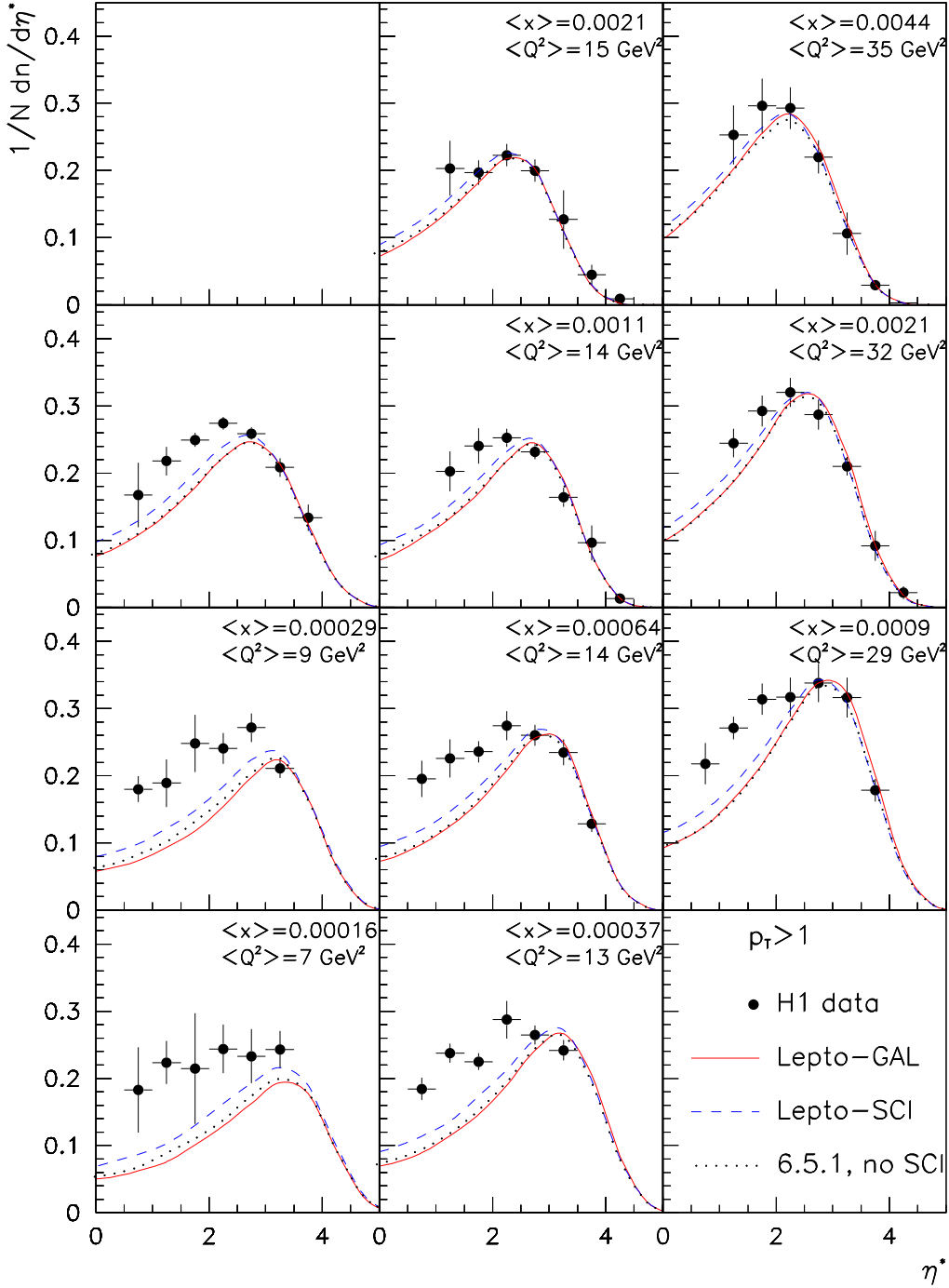,width=14cm}}
\end{center}
\vspace*{-10mm}
\caption[*]{{\it
Pseudo-rapidity distribution of charged particles  with $p_T>1$ GeV in the
hadronic cms of inclusive DIS events.
  }}
\label{fig:etaspec}
\end{figure}

\begin{figure}[t]
\begin{center}
\mbox{\epsfig{figure=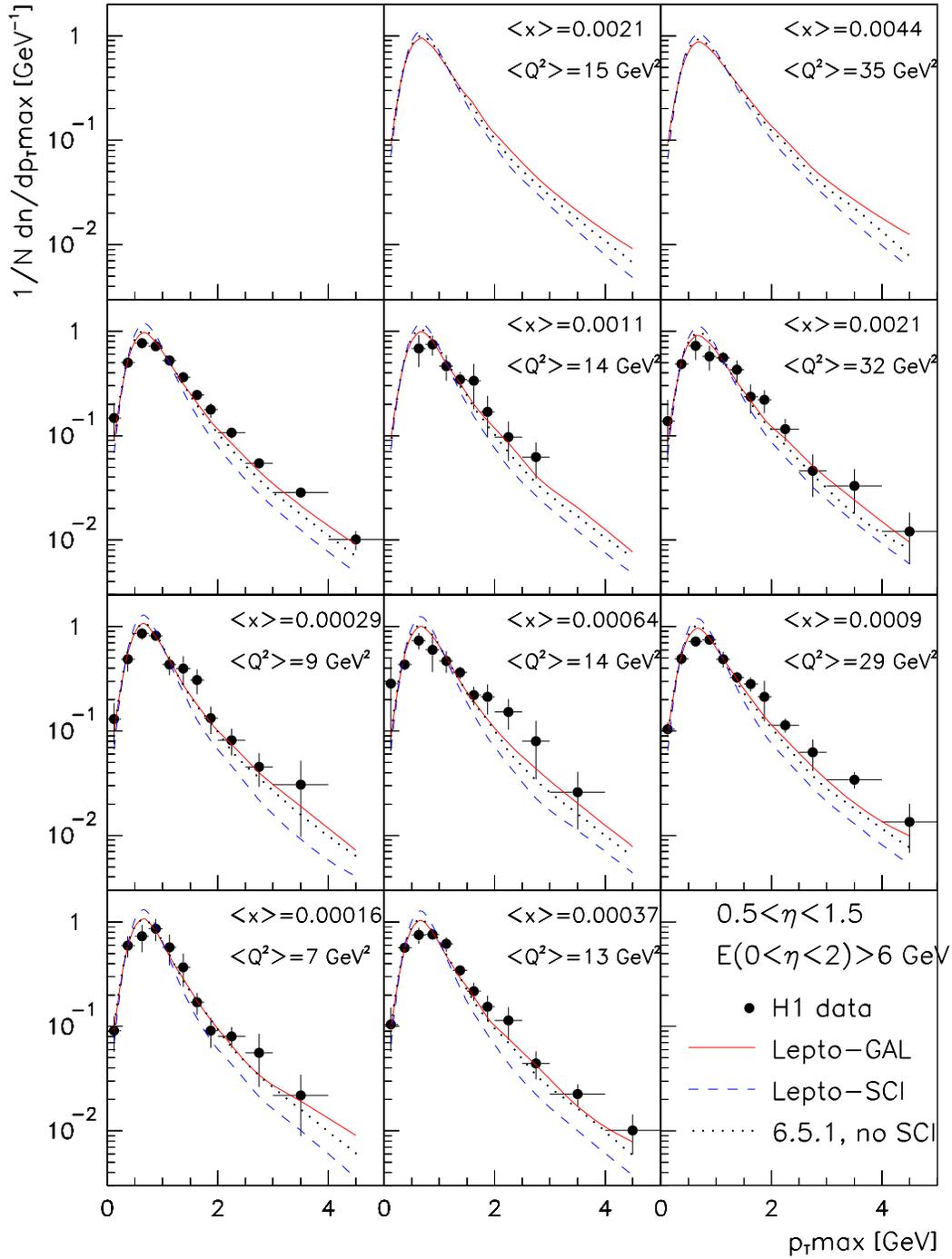,width=14cm}}
\end{center}
\vspace*{-5mm}
\caption[*]{{\it
The $p_\perp$ distribution in inclusive DIS events with large energy in central 
region, $E(0<\eta<2)>6$ GeV.
  }}
\label{fig:ptmaxspec}
\end{figure}

\newpage
Looking at the pseudo-rapidity distribution of charged particles  with
$p_\perp$ larger than 1 GeV changes the picture as shown in 
Fig.~\ref{fig:etaspec}. Now both models as well as {\sc Lepto} without string
reconnections give too few particles in the central region. Thus one should not
expect either version of {\sc Lepto} to give the correct average transverse
energy flow unless the lack of high-$p_\perp$ particles is compensated by too
many soft ones.  From this one might be tempted to draw the conclusion that the
cascade in {\sc Lepto} gives the wrong $p_\perp$ distribution.  However, this
need not be the case.  The $p_\perp$ distribution in Fig.~\ref{fig:ptmaxspec}
for events with large  energy in the central region is well described by the
GAL model and essentially also by {\sc Lepto} without reconnections.  Thus the
$p_\perp$ distribution is well reproduced by the cascade but there are too few
events with large energy in the forward region.  For the SCI model, on the
other hand, more forward energy is made up of soft particles from `zig-zag'
shaped strings resulting in a too soft  $p_\perp$ distribution. 

\begin{figure}[t]
\begin{center}
\mbox{\epsfig{figure=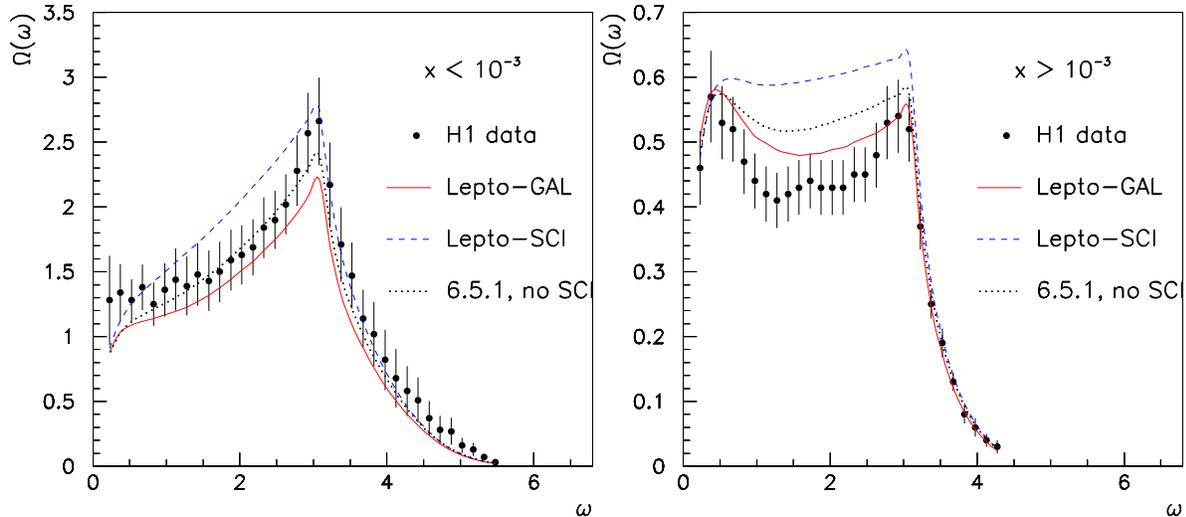,width=16cm}}
\end{center}
\caption[*]{{\it
Transverse energy-energy correlation in inclusive DIS events. 
H1 data \cite{H1-eecorr} compared to the models; GAL (full), SCI (dashed)
and no string reconnections (dotted).  
  }}
\label{fig:omega}
\end{figure}

Another instructive observable is the energy-energy correlation which in 
$e^+e^-$ annihilation has been useful to study the internal structure of jets. 
In DIS one defines \cite{BIS} the transverse energy-energy correlation 
$\Omega(\omega)=1/N_{event}\sum_{events}\sum_{i\ne j} E_{\perp i}E_{\perp j}/
Q^2(1-y)$  between pairs ($ij$) of hadrons separated a distance 
$\omega_{ij}=\sqrt{(\eta_i -\eta_j)^2+(\phi_i -\phi_j)^2}$. 
Fig.~\ref{fig:omega} shows this correlation in the two models and without
reconnections compared to data from H1 \cite{H1-eecorr}.  The SCI model has the
wrong shape since the correlation is smeared out  due to the formation of
`zig-zag' shaped strings. The suppression of such  `long' strings in GAL avoids
this and produces a reasonably good description  of the data. 

\section{Summary and conclusions}
We have shown that both the SCI and GAL models give satisfactory descriptions 
of the diffractive structure function and of more detailed hadronic properties 
of the $X$-system such as the energy flow and the seagull plot. However, when
comparing with detailed properties of inclusive DIS final states it is clear
that the SCI model fails in some respects, whereas the GAL model gives a 
description which is as good as or better than {\sc Lepto} without string 
reconnections. Specifically, the SCI model gives too many soft particles both
in current  and target regions in the Breit frame whereas the GAL model gives a
good  description of soft particles but has too few particles with large
$p_\perp$, just as when having no reconnections, which results in the average
transverse energy flow being too low compared to data~\cite{H1-eflow}.  At the
same time the GAL model gives a reasonable description of the
$p_\perp$-distribution in events with large  energy in the central region. Thus
it is too few events with high-$p_\perp$  emissions that is the problem and not
the modelling of the fragmentation process. 
In other words it is the cross-section for hard emissions that is too small in 
the model. This may be partly cured by adding resolved photon contributions
as in {\sc Rapgap} \cite{Rapgap}. From the energy-energy correlations it is also clear that the SCI
model  smears out the energy-energy correlations by making the string go
`zig-zag', whereas GAL only has minor effects on the energy-energy correlation.

One may consider whether the shortcomings of the SCI model are genuine or  can
be tuned away. The problems of giving too many soft particles is related to
events where the string after SCI goes back-and-forth producing a zig-zag
shape, \ie a longer string. Hadronisation will then produce more, but softer 
hadrons. This helps to reproduce the inclusive transverse energy flow 
\cite{SCI,H1-eflow}, but make the agreement with some of the above observables
worse. In principle one may be able to tune the hadronisation parameters to
recover a good description of the data. We have chosen not to attempt this,
since that would be against the principle of having a universal hadronisation
model,  with the same parameter values in DIS and $e^+e^-$.  A possible way out
for the SCI model could be to think of it not as  interactions with a
background field, but taking place generally between all partons in any type of
event. Then it should also apply to $e^+e^-$ annihilation and the modified
string topologies would require a retuning of the hadronisation parameters in
{\sc Jetset} in order to fit data. Although this might improve the ability of
the SCI model to describe DIS data, we have not embarked on such  a road
because it has no substantial theoretical justification. 
Another possibility would be to extend the SCI model with some dynamics 
that suppresses the probability to get longer strings, similarly to
the GAL model. 

The problem of too many soft hadrons is solved in the GAL model by suppressing
the probability for long and thereby `zig-zag' strings. At the same time  the
problem with too few particles with $p_{\perp}>1$ GeV   remains
and thus the   average transverse energy flow is below the data
\cite{H1-eflow}. However,  as already mentioned, the source of this problem is
to be found in the matrix elements and parton showers describing the hard
interactions and   not in the soft hadronisation model. 

In conclusion, it is far from easy to construct a single Monte Carlo model, 
based on reasonably physics input and few parameters, that can well describe
all kinds of hadronic final states in all interactions. Nevertheless, this 
should be the goal.

\end{document}